\documentclass{llncs}
\usepackage{graphicx}
\begin{document}

\title{A Survey on Visual Query Systems in the Web Era (extended version)
\thanks{The author would like to thank Rafael Bello
for making the initial collection of papers for this review.}
}

\author{Jorge Lloret-Gazo}

\authorrunning{Jorge Lloret}   % abbreviated author list (for running head)

 %%modified list of authors for the TOC (add the affiliations)
\tocauthor{Jorge
Lloret-Gazo(Universidad de Zaragoza)}

\institute{Dpto. de Inform\'atica e Ingenier\'{\i}a de Sistemas.\\
Facultad de Ciencias. Edificio de Matem\'aticas.\\ Universidad de
Zaragoza. 50009 Zaragoza. Spain.\\ \email{jlloret@unizar.es}}
\maketitle   

\begin{abstract}
As more and more collections of data are becoming available on the web to everyone,
non expert users demand easy ways to retrieve data from these collections.
One solution is the so called Visual Query Systems (VQS) where queries are
represented visually and users do not have to understand query languages such as SQL or XQuery.
In 1996, a paper by Catarci reviewed the Visual Query Systems available until that year.
In this paper, we review VQSs from 1997 until now and try to determine whether
they have been the solution for non expert users. The short answer is no because
very few systems have in fact been used in real environments or as commercial tools.
We have also gathered basic features of VQSs such as the
visual representation adopted to present the reality of interest or the 
visual representation adopted to express queries.
\end{abstract}

\section{Introduction}

In recent years, and mainly because of the arrival of the web, more and more collections of data are becoming available 
to everyone in fields ranging from biology to economy or geography. One of the consequences of this fact is that end users, 
but not experts in Computer Science, demand easy ways to retrieve data from these collections. 
 
Beginning in 1975 with Query By Example (QBE)~\cite{zloof1975query} there have been many proposals in this direction,
that is, to facilitate the work of the final user. In~\cite{DBLP:dblp_journals/vlc/CatarciCLB97}, the
authors reviewed the so-called Visual Query Systems (VQS) from 1975 to 1996
defined as

\begin{quote}
``systems for querying databases that use a visual representation to
depict the domain of interest and express related requests".
\end{quote}

In this paper, we extend the review from 1997 to date, concentrating our
efforts on visual queries to structured information, for example,
queries to underlying relational or XML databases. We do not consider the typical
search on semistructured documents such as web pages through search engines like
Google. Although they are also a good solution for end-users, in this survey we do not
take into account natural language interfaces for database query formulation.

The main goal of this survey is to answer the following question:

to what extent have the VQS been the solution for novel users for querying databases?

To answer this question, we have studied two features: web availability of
and validatioin undergone by the systems. The first feature indicates
that the system was designed to be reached easily by novel users
simply by means of a web browser, without the burden of
installation and with universal availability. The second feature
indicates the widespread use of VQSs  in practice. Thus, the more systems 
commercially available, the greater the extension reached by VQSs.

The short answer to the question is that, as far as we know, there is only
one system commercially available and designed for the web: 
Polaris ~\cite{stolte2008polaris}.

Moreover, we have included two basic features
extracted from the paper~\cite{DBLP:dblp_journals/vlc/CatarciCLB97}: 
the visual representation adopted
to present the reality of interest and the visual representation adopted to
express queries. 
With respect to web features, we have also considered
relevant whether the prototype deals with data formatted for the web, that is,
XML data or RDF data.

The rest of the paper is organized as follows. In Section 2 we state
the method followed for elaborating the survey and we briefly describe
the values of the relevant features included in the paper.
In Section 3, we have made a systematic review of the selected features
for each paper of the survey. Finally, in Section 4, we have drawn several
conclusions about the VQSs.

\section{Statement of the method}

A survey about a particular object must determine the relevant features
of the object with respect to a particular purpose. Once the features have been determined, 
the next step is to find the possible values of these 
features. Finally, we have to determine the best combinations of 
the pairs (feature, value) for the particular purpose.

Usually, we can extract the relevant features and their possible values
from published papers about the object, by assuming features in their entirety
or by adapting them to new perspectives appearing after the papers 
have been published. Moreover, we can add features detected by ourselves
which were not previously included in any paper.

The survey develops through several steps, which are usually interspersed.
In the first step, a complete search of sources determines the candidate
papers that deal with the object. In the second step, the relevant
features of the object with respect to the particular purpose are determined.

Our object in this survey are the visual query systems
with the purpose of facilitating querying databases to non expert in Computer
Science users. 

The survey~\cite{DBLP:dblp_journals/vlc/CatarciCLB97} reviews up to 80 references
from 1975 until 1996 used for querying
traditional databases. For this survey, we have searched for papers related with VQS from 1997 to date and
we have found 194 candidate papers. Next, we have discarded papers about query languages but
without visual part (122) and papers about natural language query languages(8)
because they deserve a separate survey. In the remaining 64 works, we have determined
sets of `similar papers' and we have discarded all but one paper in each set.
A set of similar papers is composed of several papers built on different aspects 
of the same idea for a VQS. They also include preliminary versions of the VQS which
were later on subsumed by more complete journal publications. We have found
30 similar papers. So, we have discarded 122 + 8 + 30 papers, that is, 160 papers.
As a result, the number of papers reviewed in this survey is
34.
In Table~\ref{tab:numbersummary} we summarise all these figures.

\begin{table}%
\caption{Summary of paper figures}
\label{tab:numbersummary}
\begin{tabular}{|p{9cm}|p{1cm}|}
\hline
& Number\\\hline
Candidate papers & 194\\\hline
excluded because they do not include a visual part& -122
\\\hline
excluded because they deal with natural language query& -8
\\\hline
excluded becacuse they come from sets of similar papers& -30
\\\hline
Total papers included in the survey: & 34
\\\hline

\end{tabular}
%%}
\end{table}%

As for relevant features, we have extracted the following
from the survey of Catarci~\cite{DBLP:dblp_journals/vlc/CatarciCLB97}:

\begin{enumerate}
\item{Visual representation adopted to present the reality of interest}
\item{Visual representation adopted to express queries}
\end{enumerate}

The values of this feature have been determined from the work
~\cite{DBLP:dblp_journals/vlc/CatarciCLB97} and from other papers,
such as~\cite{DBLP:dblp_conf/jcdl/ClarksonNF09}, where the faceted option appeared.

For answering the question of this paper, we have added the following features

\begin{enumerate}
\item{Web orientation}
\item{Validation}
\end{enumerate}

Let us explain briefly each of the features as well as their values.

\subsection{Visual representation adopted to present the reality of interest}
This feature has been borrowed from the work of Catarci~\cite{DBLP:dblp_journals/vlc/CatarciCLB97}.
The reality of interest is modeled by a designer by means of a data metamodel as,
for example, the entity/relationship metamodel or a graph data metamodel. As a result of the
modelization process, a data model is obtained and it is presented to the
user so that (s)he formulates queries on it. The ways the data
model is presented to the user are:

\begin{itemize}
\item{Diagram-based}
\item{Icon-based}
\item{Form-based}
\item{Faceted}
\item{Unknown}
\end{itemize}

The category in which each selected paper falls as well as a more detailed explanation
of some of the papers are given in Section~\ref{systematic}.

\subsubsection{Diagram-based} 
Data metamodels come with an associated typical representation
for their elements. For example, in the entity/relationship metamodel, there are
many representations available and one of them consists of drawing
rectangles for the entity types, diamonds for the relationship types and ovals for
the attributes. In the diagram-based option, the user has available a diagrammatical
representation of the data model elaborated with the typical graphical 
representation for the elements of the metamodel.

\subsubsection{Icon-based} Unlike the diagram-based approach, in this representation
there are only iconic representations of some elements of the data model, but the
user does not have available the complete data model.
According to Catarci~\cite{DBLP:dblp_journals/vlc/CatarciCLB97}, `these VQS are mainly addressed
to users who are not familiar with the concepts of data models and may find it difficult
to interpret even an E-R diagram'. The aim of the icons is to represent a certain concept
by means of its metaphorical power. The problem of these systems is how to construct 
them in such a way that they express a meaning which is understandable without ambiguity
to the users.

\subsubsection{Form-based} The typical forms of web pages serve for presenting
the extensional database. This occurs in papers such as~\cite{stolte2008polaris}.

\subsubsection{Faceted}  The data are modeled as faceted classifications
which organize a set of items into multiple, independent taxonomies. Each classification
is known as a facet and the collection of classification data is faceted metadata. The specific
category labels within a facet are facet values. 

For example, the set of items can be architectural works. For these items, 
the facets are the architect, the location or the materials. The facet values
for materials are stone, steel, etc

\subsubsection{Unknown} As the data model always exists, this option refers to the case
where the data model is unknown. 
For example, the data model may be presented
in a paper in textual form but there is no explanation about the way it
is presented to the user. 
For example, paper~\cite{DBLP:journals/cj/Owei03} hides the database and tries to guess the paths for
the query from the entities chosen by the user.

\subsection{Visual representation adopted to express the queries}
This feature has been borrowed from the work of Catarci~\cite{DBLP:dblp_journals/vlc/CatarciCLB97} and
we have adapted it to the object of the survey by adding the Faceted value. The list of values is:

\begin{itemize}
\item{Diagram-based}
\item{Icon-based}
\item{Form-based}
\item{Faceted}
\end{itemize}

Next, we describe briefly the values of this feature.

\subsubsection{Diagram-based}

The diagram-based option means that the query is expressed on a diagrammatic
representation of the data model.

\subsubsection{Icon-based}
The icon-based option includes two cases. 
In the first case, the system offers icons for representing the elements involved in the query.
For building a query, the user drags and drops the appropriate icons into a canvas.
The second case is
the same as in~\cite{DBLP:dblp_journals/vlc/CatarciCLB97}, where the icons
`denote both the entities of the real world and the available functions of the system'.

\subsubsection{Form-based}
Another way to facilitate the query is the form option where the user composes
the query by completing options of different elements of a form. 
The drawback is that the query logic of the end-user does not always fit into a form.
%%We have found
%%this situation only in the papers~\cite{stolte2008polaris}  and~\cite{DBLP:journals/cj/Owei03}.

\subsubsection{Faceted}
We have added as a new value `Faceted' for describing 
a system which includes 
data and metadata  in the same page. There, the user specifies the query by
clicking on the appropriate links. We have found
this situation only in one paper~\cite{DBLP:dblp_conf/jcdl/ClarksonNF09}.

\subsection{Web orientation}

For the web orientation, we have selected two features which are not
mutually orthogonal.
The first feature is whether
the prototype is working on the web or has been conceived to be used
in local mode. For the first situation, the value is {\em Available on the web}
and this means that the final user can query the database by means of
a prototype which is working on the web. The two values are:

\begin{itemize}
\item{There is no web orientation}
\item{Available on the web}
\end{itemize}

The second feature indicates whether the user can query data formatted
for the web and the values are:

\begin{itemize}
\item{Data not formatted for the web}
\item{Query XML data}
\item{Query RDF data}
\end{itemize}

The values are not orthogonal. So, a paper can have the two values.
This is the case, for example, of paper~\cite{DBLP:journals/tods/BragaCC05}.

\subsection{Validation}

The validation of an idea can be done from several points of view.
Regarding query systems, there are, at least, two dimensions:
usability and performance.

For example, paper~\cite{choi2014vxq} focuses on performance
and explains query rewriting techniques that improve
the query evaluation performance so that the query exection time
is reduced.

However, in this paper we concentrate on the usability dimension,
that is, the experiments made with users in order to determine
the ease of use of the proposed prototype.
For this feature, the list of values is:

\begin{itemize}
\item{Only prototype}
\item{Prototype tested with users}
\item{Prototype tested in a real environment}
\item{Commercial tool}
\end{itemize}

Next, we describe briefly each value of this feature.
The option {\em only prototype} means that a prototype has been
built but no test has been made with users. The value
{\em prototype tested with users} means that several experiments
have been carried out in order to determine the usability of the 
prototype. The value {\em prototype tested in a real environment}
means that it has been used for real tasks in a particular setting,
for example in a department of a university. Finally, the option
{\em commercial tool} means that the VQS has been fully implemented,
offered to the public and is in real use in diverse installations.

\subsection{Other features}

We were initially also interested in the expressive power of the query
languages of the revised papers. Here, the difficulty is that many
papers do not offer the complete specification and only deal with
query examples. For this reason, we had to discard this feature.

As a consequence, we also had to give up to offer in this review
the same queries for all the visual query systems because with
the explanation of the papers we were not able to specify the queries
in all the visual query systems.

\begin{table}%
\caption{Visual Query Systems(1997-2003)}
\label{tab:features1}
\begin{tabular}{|p{0.8cm}|p{1.5cm}|p{1.3cm}|p{2.2cm}|p{4.4cm}|}
\hline
{\bf Cite} & {\bf Database} & {\bf Query} & {\bf Web} & {\bf Validation}\\\hline
\cite{DBLP:conf/icde/BalkirSOO96} & Unknown & Icon & No & Only  prototype\\\hline
\cite{sengupta1997query} & Unknown & Form & No & Tested with users\\\hline
\cite{DBLP:dblp_conf/er/BloeschH97} & Diagram & Diagram & No & Only prototype\\\hline
\cite{DBLP:dblp_journals/vldb/CatarciSC97} & Diagram & Diagram & No & Tested with users\\\hline
\cite{DBLP:conf/ideas/MadurapperumaGF97} & Diagram & Diagram & No & Only prototype\\\hline
\cite{DBLP:dblp_conf/avi/ShinGC98} & Diagram & Diagram & No & Only prototype\\\hline
\cite{DBLP:dblp_conf/avi/MurrayPG98} & Icon & Icon & No & Only prototype\\\hline
\cite{DBLP:dblp_journals/vlc/BenziMR99} & Diagram & Diagram & No & Tested with users\\\hline
\cite{cohen1999equix} &Unknown& Form& Query XML data & No \\\hline
\cite{DBLP:dblp_journals/vlc/CruzL01} & Icon & Icon & Available on the web & Tested with users\\\hline
\cite{DBLP:journals/tkde/PoulovassilisH01} & Diagram & Diagram & No &  Only  prototype\\\hline
\cite{DBLP:dblp_journals/is/SilvaCS02} & Icon & Icon & No &  Tested with users\\\hline
\cite{papakonstantinou2002qursed} &Diagram & Form & Query XML data& Only prototype\\\hline
\cite{DBLP:dblp_journals/vlc/NarayananS02} & Icon & Icon & No & Tested with users\\\hline
\cite{Morris:2002:VQL:1556262.1556321} & Unknown & Icon & No & Only prototype\\\hline
\cite{berger2003xcerpt} & Unknown & Icon & Query XML data & Only prototype\\\hline
\cite{DBLP:journals/cj/Owei03} & Form & Form & No & Tested with users\\\hline
\cite{abraham2003foxq,erwig2003xing} & Unknown & Form & Query XML data & Only prototype\\\hline

\end{tabular}
%%}
\end{table}%

\begin{table}[t]
\caption{Visual Query Systems(2004-2014)}
\label{tab:features2}
\begin{tabular}{|p{0.8cm}|p{1.5cm}|p{1.3cm}|p{2.2cm}|p{4.4cm}|}
\hline
{\bf Cite} & {\bf Database} & {\bf Query} & {\bf Web} & {\bf Validation}\\\hline
\cite{DBLP:journals/tods/BragaCC05} & Diagram & Icon & Available on the web; Query XML data & Only prototype\\\hline
\cite{meuss2005visual} & Diagram & Diagram & Query XML data & Tested with users\\\hline
\cite{DBLP:dblp_conf/www/HarthKD06} & Unknown & Icon & Query RDF data & No\\\hline
\cite{DBLP:dblp_conf/avi/RontuKM06} & Diagram & Diagram & No & Tested in a real environment\\\hline
\cite{DBLP:dblp_conf/sigmod/JagadishCEJLNY07} & Unknown & Form  & Query XML data & Tested in a real environment \\\hline
\cite{DBLP:dblp_journals/dke/TerwilligerDL07} & Unknown & Form & No & Tested in a real environment \\\hline
\cite{DBLP:dblp_conf/sac/SansL08} & Diagram & Icon & Query XML data & Only prototype\\\hline
\cite{stolte2008polaris} & Form & Form &  Available on the web & Commercial tool\\\hline
\cite{DBLP:dblp_conf/jcdl/ClarksonNF09} & Diagram & Faceted & Available on the web & Only prototype  \\\hline
\cite{DBLP:dblp_journals/internet/JarrarD10} & Unknown & Diagram & Query RDF data & Only prototype\\\hline
\cite{varga2010conceptual} & Diagram & Icon & No & No\\\hline
\cite{borges2010feasible} & Diagram & Diagram & Available on the web & Tested with users\\\hline
\cite{jin2010gblender} &Unknown & Diagram & No & Tested with users\\\hline
\cite{storrle2011vmql} & Diagram  & Icon & No & Tested with users\\\hline
\cite{wu2012gmql} &Unknown & Icon &  No & No \\\hline
\cite{choi2014vxq} & Diagram  & Diagram & Query XML data& Only prototype\\\hline
\end{tabular}
%%}
\end{table}%

\section{Systematic review}
\label{systematic}
In this Section, for each feature, we include comments only about a portion of the papers
gathered in Table~\ref{tab:features1} and in Table~\ref{tab:features2}. The reason is that
we only comment on papers of special relevance with regard to the intended feature.
For the rest of the papers, only the value of the feature in Table~\ref{tab:features1} and Table~\ref{tab:features2}
is available.

%%%INICIO REPRESENTACIÓN DE LA REALIDAD
\subsection{Visual representation adopted to present the reality of interest}

In column 2 of Table~\ref{tab:features1} and Table~\ref{tab:features2},  we summarise the options of the VQSs for specifying
the database.

\subsubsection{Diagram-based}

The queries in the Conquer-II query language~\cite{DBLP:dblp_conf/er/BloeschH97}
are written in a conceptual schema based on the ORM conceptual modeling approach.
In ORM, the world is modelled in terms of objects and roles and the notion of
attribute is not used.

Paper~\cite{DBLP:dblp_journals/vldb/CatarciSC97} deals with heterogeneous
databases. It provides a formally defined and semantically rich data model, the
Graph Model.
%%and a minimal set of Graphical Primitives,
%%in terms of which general query operations may be visually
%%expressed. 
A conceptually single database, the Graph Model Database,
is built for the different heterogeneous databases and the user
formulates his/her queries against this model.

Paper ~\cite{DBLP:conf/ideas/MadurapperumaGF97} adopts the EER model to present
the reality of interest. It presents the MiTRA system, composed of several modules. One of them,
the Schema Visualization System, extracts database schema information and
presents the visual structure on the user's screen in the form of an EER schema in
the schema visualisation window, which is always available.

Paper~\cite{DBLP:dblp_conf/avi/ShinGC98} proposes the EPISQUE interface, where
the epistemic diagrams represents the real world in terms of entities, events
and states. It has the idea that ER diagrams do not provide enough
information so that people outside the organization can form SQL queries against the
database. So, the E/R diagrams are extended to epistemic diagrams.

%%In~\cite{DBLP:dblp_conf/ssdbm/SilvaC99}  the database
%%schema is visually represented in three different ways: as
%%a top-down tree, as a graph and as a subgraph. Through 
%%this representation, the user has a global view of the classes of the
%%schema and their interrelationships.

\begin{figure}
\centerline{\includegraphics[width=40mm]{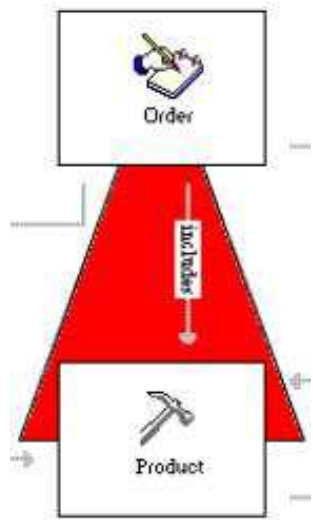}}
\caption{A fragment of a vision in VISIONARY.}
\label{fig:vision}
\end{figure}

In~\cite{DBLP:dblp_journals/vlc/BenziMR99}, the model of the real world is called vision.
A vision is composed of visual concepts and associations between them and
is a clear representation of a relational database. Several visions may be built on the
same database, to deal with different kinds of users. An example of a vision
can be seen in Figure~\ref{fig:vision}. It is a combination of the diagram and of the icon
approaches.

\begin{figure}
\centerline{\includegraphics[width=60mm]{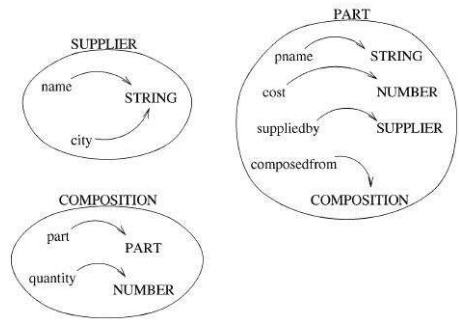}}
\caption{A fragment of a schema in Hyperlog.}
\label{fig:hyperlog}
\end{figure}

The data model in Hyperlog of~\cite{DBLP:journals/tkde/PoulovassilisH01} is called
the Hypernode Model. The schema of the database is a set of hypernode types,
which may be recursively defined. An example of schema in Hyperlog can be seen in
Figure~\ref{fig:hyperlog}.
%%detallar esto

Paper~\cite{DBLP:journals/tods/BragaCC05} presents XQBE, a tool
for querying XML data. The XML schema or DTD specification
is offered to the user as a starting point 
for specifying queries in a navigational way.

\begin{figure}
\centerline{\includegraphics[width=60mm]{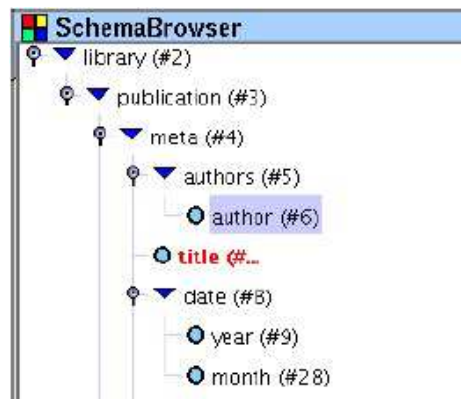}}
\caption{Schema in $X^{2}$}
\label{fig:meuss1}
\end{figure}

In paper~\cite{meuss2005visual} the XML structure is visualized as a GUI tree
on the left part of the SchemaBrowser. There, nodes can be
activated and can serve as a basis for automatically generating a query tree.
An example of a library schema is shown in Figure~\ref{fig:meuss1}.

In paper~\cite{DBLP:dblp_conf/avi/RontuKM06} 
the data model consists of classes and their attributes
as well as associations between classes. In the interface,
there are three canvases and the application model is displayed
in the model canvas. The user can browse the data model 
to find the right concepts for the query.

In~\cite{DBLP:dblp_conf/sac/SansL08}, an interface for ordered XQuery, called IFOX, is described.
In this approach, the schema is visually presented as a tree, where 
the elements, the attributes and the terminal elements are clearly distinguished.
The XML schema is specified by the user or is built by a
schema constructor.

\begin{figure}
\centerline{\includegraphics[width=60mm]{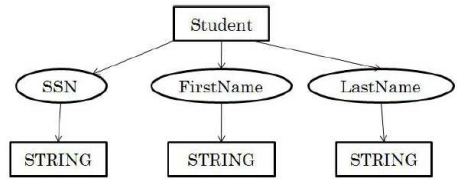}}
\caption{A fragment of a schema in CGBDI.}
\label{fig:cgbdi}
\end{figure}

In paper~\cite{varga2010conceptual}, the database schema is specified as
a unified conceptual graph. For example, the student table is represented as in
Figure~\ref{fig:cgbdi}.

In~\cite{borges2010feasible}, queries are expressed on data warehouse schemas composed
of tables. A typical schema consists of a fact table and several dimension tables related
with the fact table. This type of structure is commonly known as a star schema.

\begin{figure}
\centerline{\includegraphics[width=60mm]{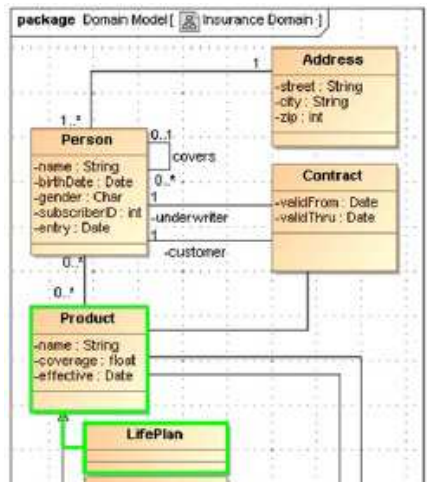}}
\caption{A fragment of data in VMQL.}
\label{fig:vmql}
\end{figure}

Paper~\cite{storrle2011vmql}  presents VMQL, a language for querying models
used in large scale model based development. VMQL uses 
the respective modeling language of the source model as the query language.
So, if the source model is written in UML, the queries are also written using UML.
In Figure~\ref{fig:vmql}, we can see an example for data, in this case, a UML
class diagram.

Paper~\cite{choi2014vxq} offers an interface divided into two parts. In the left part,
a diagram of an XML document is shown 
to represent source XML documents, lets users explore the documents, 
and selects parts of the documents to be used in queries.

\subsubsection{Icon-based}

Paper~\cite{DBLP:dblp_conf/avi/MurrayPG98} shows classes and extents.
For a class, the only visible information is its  icon and its name.

%%In Delaunay,~\cite{DBLP:dblp_journals/vlc/CruzL01}, users query visually
%%an object-oriented database. The classes are listed in the Delaunay tool.

In~\cite{DBLP:dblp_journals/vlc/NarayananS02} the queries are built on the relational
model. Queries are built on the Vis\_table, which is a logical view of a relation 
in the database and may be a join of more than one physical table. The attributes
of this table are available as icons in the interface for building the iconic queries.

Paper~\cite{DBLP:dblp_journals/is/SilvaCS02}  presents a Temporal
Visual Query Environment(TVQE) for historical databases. In this environment,
the data model is called Temporal Graph Model(TGM) and has a set of Temporal
Graphical Primitives(TGPs). The schema includes classes, attributes, 
relationships between classes, temporal classes, temporal attributes and temporal relationships.
The database schema
adopts an iconic representation which exploits a graphical
notebook metaphor.

\subsubsection{Form-based}
Paper~\cite{stolte2008polaris} presents the schema in a form-based manner. The user
drags the fields of a query into shelves. Polaris interacts with multidimensional relational
databases in the form of spreadsheets or available inside the DBMS.

\subsubsection{Faceted}
In paper~\cite{DBLP:dblp_conf/jcdl/ClarksonNF09}, the data are modeled as faceted classifications
which organizes a set of items into multiple, independent taxonomies. Each classification
is known as facet and the collection of classification data is faceted metadata. The specific
category labels within a facet are facet values. 

For example, the set of items can be architectural works. For these items, 
the facets are the architect, the location or the materials. The facet values
for materials are stone, steel, etc.

\subsubsection{Unknown}

The VISUAL system of~\cite{DBLP:conf/icde/BalkirSOO96} is a query language intended
for handling scientific experiment data. It offers a textual representation 
of the object-oriented data model on which the query system
is built. The data model includes classes as Frame or Particle.

In paper~\cite{DBLP:journals/cj/Owei03}, the schema is not directly presented to the user.
Basically, the user specifies the beginning and the end of the query and CQL uses 
built-in meta-knowledge about the application schema to find the complete path.
For expert users, the interface has the option of seeing the complete 
database application diagram. This is a semantically constrained entity-relationship diagram
which includes entity types and associations among themselves.

Paper~\cite{DBLP:dblp_conf/sigmod/JagadishCEJLNY07} proposes the MiMi system, a deep integration
of several protein interaction databases. The entire dataset and metadata were stored in Timber, a 
native XML database but it is not clear how the schema is made available to the users.

Paper~\cite{DBLP:dblp_conf/www/HarthKD06} deals with web query languages but 
does not offer a schema since semistructured data typically comes without a fixed schema.

%%%FIN REPRESENTACIÓN DE LA REALIDAD

%%VISUAL REPRESENTATION ADOPT TO EXPRESS THE QUERIES
\subsection{Visual representation adopted to express the queries}
In column 3 of Table~\ref{tab:features1} and Table~\ref{tab:features2},  we summarise the options of the VQSs for 
formulating the query.

\subsubsection{Diagram-based}

\begin{figure}
\centerline{\includegraphics[width=90mm]{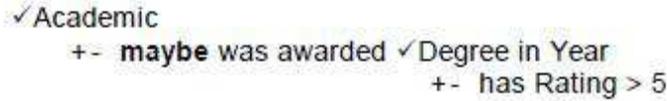}}
\caption{Example of query  in Conquer-II.}
\label{fig:conquer}
\end{figure}

Conquer-II queries~\cite{DBLP:dblp_conf/er/BloeschH97} are mapped into SQL queries
for execution in commercial DBMS. The queries are shown in textual form as
a tree of predicates connecting objects. For example, a query like 
`each academic as well as
their degrees (if any) that are worth more than a 5 rating' is expressed 
as shown in Figure~\ref{fig:conquer}.

The user specifies a query by dragging to the query pane an object type from an
object pick list. Then, another pane displays the roles played by that object
and the user selects the appropriate ones. In this way, users specify queries without
knowing the underlying structure.

\begin{figure}
\centerline{\includegraphics[width=80mm]{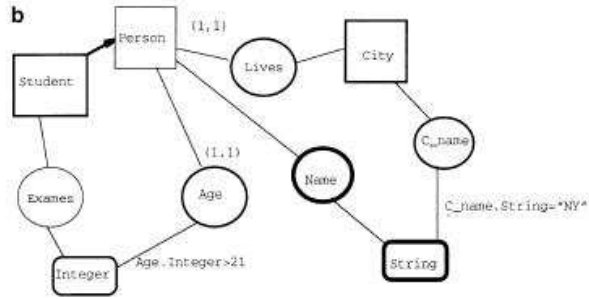}}
\caption{Example of query formulation for heterogeneous databases.}
\label{fig:heterogeneous}
\end{figure}

For specifying queries in the Graph Database Model of~\cite{DBLP:dblp_journals/vldb/CatarciSC97},
two graphical primitive are defined: selection of nodes and drawing of edge. 
The primitives are used to express any query-oriented interaction with the database.
For example, the query `retrieve the names of all the students whose age is greater
than 21 and living in NY' is expressed graphically as shown in Figure~\ref{fig:heterogeneous}.

%%With regard to the expressive power, it was proven in~\cite{DBLP:dblp_journals/is/CatarciSA93}
%%that `the class of queries
%%computed by the graphical primitives contains the class of queries computable
%%by the Relational Algebra'.

\begin{figure}
\centerline{\includegraphics[width=130mm]{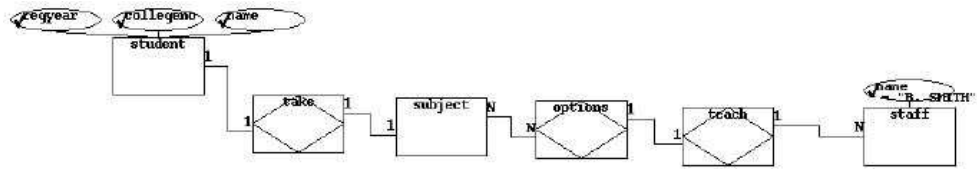}}
\caption{Example of query formulation in the MiTRA system.}
\label{fig:mitra}
\end{figure}

The queries in MiTRA~\cite{DBLP:conf/ideas/MadurapperumaGF97} are expressed on an EER schema.
For doing so, the Graphical Query System allows users to manipulate
the EER schema. It presents the available graphical query operators to the user,
allowing the formulation of queries involving selection, projection, join and recursion.
For example, 
the query `Find the names of all students taught by R. SMITH.' 
is specified as shown in Figure~\ref{fig:mitra}.

In EPISQUE~\cite{DBLP:dblp_conf/avi/ShinGC98}, the user creates queries in several steps. 
First, (s)he places restriction, projection and join on epistemic diagrams. Next, they are
converted into SQL expressions and finally they are sent to the targeted
db for processing.

For the specification of queries in VISIONARY of ~\cite{DBLP:dblp_journals/vlc/BenziMR99},
the notion of viewpoint is used. The expressive power of VISIONARY is limited to
restricted join queries enriched with aggregation functions. The visual queries are translated
into SQL and then to the DBMS. Basically, a viewpoint is a primary concept of the vision 
together with  several associations starting from the primary concept. 
The formulation of a query consists of the following steps:
(1) choose a primary concept, (2) edit the viewpoint, 
(3) choose attributes to be retrieved,
(4) formulate selections on attributes and
(5) order and/or group the results.
%%This specifcation is hybrid because it combines diagrams and icons.

\begin{figure}
\centerline{\includegraphics[width=60mm]{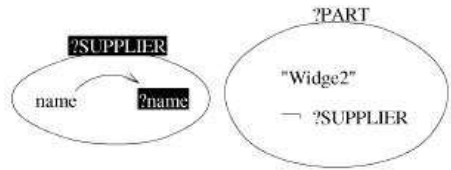}}
\caption{Query in Hyperlog}
\label{fig:queryHyperlog}
\end{figure}

Paper~\cite{DBLP:journals/tkde/PoulovassilisH01} presents
Hyperlog, a declarative, graph-based language that 
supports database querying. A Hyperlog query consists of
a number of graphs which are matched
against the hypernodes in the database and which generate
graphical output. The query 
`suppliers who do not supply Widge2' is specified in Hyperlog as in Figure~\ref{fig:queryHyperlog}.

Hyperlog admits the negation, conjunction, and
disjunction of queries.
Built-in functions and relational operators can be
visualized as hypernodes, and 
intentional information can be derived by encapsulating
programs within templates.

\begin{figure}
\centerline{\includegraphics[width=60mm]{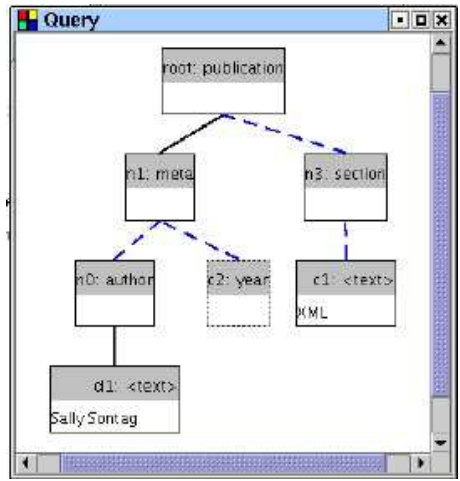}}
\caption{Query in $X^{2}$}
\label{fig:meuss2}
\end{figure}

In X2~\cite{meuss2005visual}, the user selects nodes of interest in the XML diagram and the system
automatically generates query trees from the nodes selected.
An example can be seen in Figure~\ref{fig:meuss2}
where the query `I would like to see interesting recent papers on
XML written by Sally Sonntag' is shown.

\begin{figure}
\centerline{\includegraphics[width=90mm]{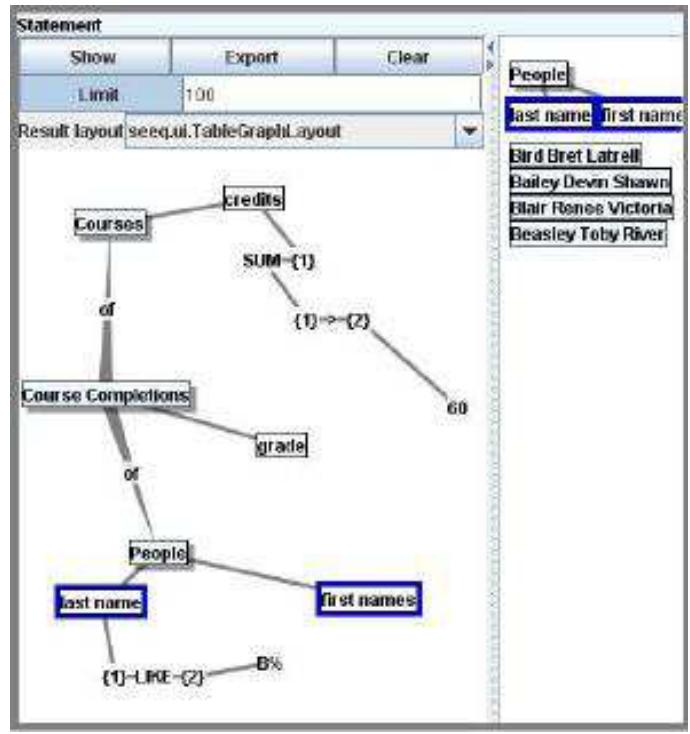}}
\caption{Query in SEEQ}
\label{fig:seeq}
\end{figure}

In SEEQ~\cite{DBLP:dblp_conf/avi/RontuKM06} queries are depicted
in the query canvas. They can contain operators, constants and sub-queries.
For example, the query `find the students
whose last name begins with `B' and who have more than
60 credits' is depicted in Figure~\ref{fig:seeq}.

Paper~\cite{DBLP:dblp_journals/internet/JarrarD10} proposes a query-by-diagram language called MashQL.
With this language, people query the web by building data mashups diagramatically. The paper assumes
that web data sources are represented in RDF, and SPARQL is the query language to which the graphical mashup
is translated. The novelty of MashQL is that it allows querying a data source without any prior understanding 
of the schema or the structure of the source.

\begin{figure}
\centerline{\includegraphics[width=60mm]{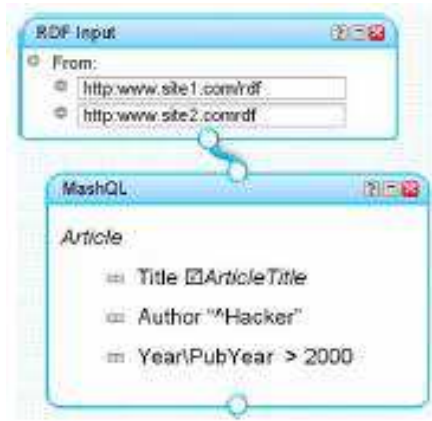}}
\caption{Example of query formulation in MashQL.}
\label{fig:mashup}
\end{figure}

The specification of mashups is as follows: After a user selects the dataset in the RDF input module,
the user selects the query subject in a drop-down list generated from the dataset. To add a restriction on the
chosen object, a list of possible properties is dynamically generated. For the selected properties, user may then
choose an object filter. The query 
`retrieve
Hacker's articles published after 2000 from two web locations'
expressed with MashQL is shown in Figure~\ref{fig:mashup}.

VISQUE~\cite{borges2010feasible} presents as a main feature 
the possibility of new queries such as union, intersection, in and not in, not covered in other papers
but there are also other complex query operations not covered in the paper. I have not found
any specific example in this paper.

\begin{figure}
\centerline{\includegraphics[width=120mm]{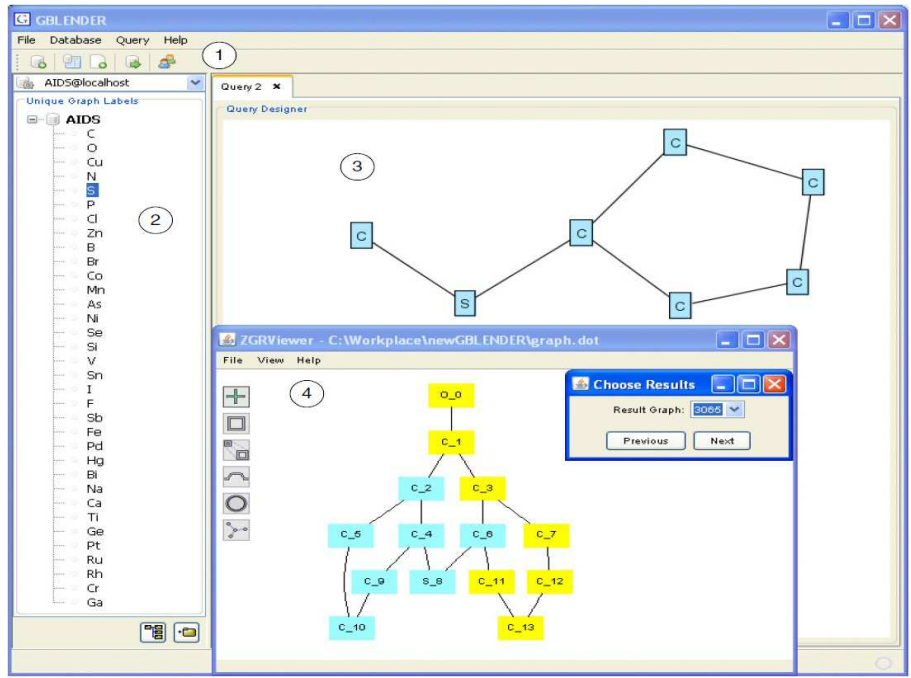}}
\caption{Example of query formulation in GBlender.}
\label{fig:gblender}
\end{figure}

GBLENDER~\cite{jin2010gblender} proposes the problem of finding those
graphs of a graph database which contains a query graph.
The user formulates a query by clicking-and-dragging items on the
query canvas. In  Figure~\ref{fig:gblender} we can see the formulation of a query
but the paper does not offer the corresponding textual version.

\begin{figure}
\centerline{\includegraphics[width=60mm]{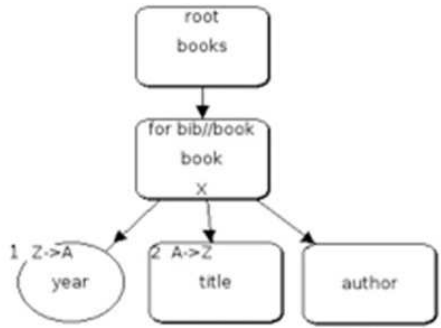}}
\caption{Example of query formulation in VXQ.}
\label{fig:vxq}
\end{figure}

VXQ~\cite{choi2014vxq} offers an interface divided into two parts. In the right part,
the user composes queries 
by exploring the diagram of the XML document displayed on the left pane of the prototype.
For example, the query 
`create a flat list of all title-author pairs with the year of publication for all books
published by Addison-Wesley after 1991, and present these books in reverse chronological 
order by years and then in alphabetical order by titles for each year'
 is specified as
can be seen in Figure~\ref{fig:vxq}.
%%Termina diagram-based

\subsubsection{Icon-based}

\begin{figure}
\centerline{\includegraphics[width=80mm]{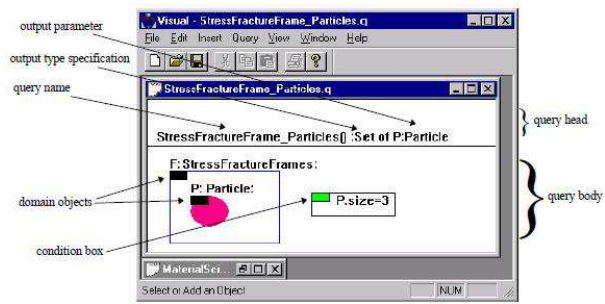}}
\caption{Example of query  in VISUAL.}
\label{fig:particle}
\end{figure}

In the VISUAL system of ~\cite{DBLP:conf/icde/BalkirSOO96}, queries are implemented as
objects and interact with each other for query processing. An example of a query is 
`find the particles each of which has a size equal to 3 
microns and resides in the set of frames named StressFractureFrames'
(see its graphical representation in Figure~\ref{fig:particle}).
The VISUAL queries are mapped into `select from where' queries in OQL.

\begin{figure}
\centerline{\includegraphics[width=70mm]{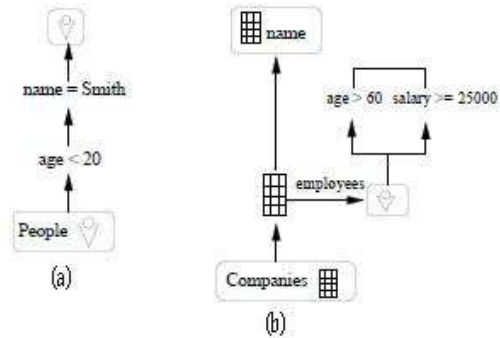}}
\caption{Example of query formulation with join in Kaleidoquery.}
\label{fig:kaleido1}
\end{figure}

Paper~\cite{DBLP:dblp_conf/avi/MurrayPG98} presents the graphical query language
Kaleidoquery, whose target language is OQL. Both languages have the same expressive
power. The queries follow the filter flow model. The input are class instances and their
information is filtered by constraints placed on the attributes of the class. For example, 
Figure~\ref{fig:kaleido1}(a) represents the query `find the people whose age is less than 20
and whose name is Smith'. The instances from the input extent flow through the query
and are filtered through the two constraints.
The query `find the name of the companies which have at least one employee who
earns more than 25000 euros and is older than 60' is represented in Figure~\ref{fig:kaleido1}(b).

\begin{figure}
\centerline{\includegraphics[width=70mm]{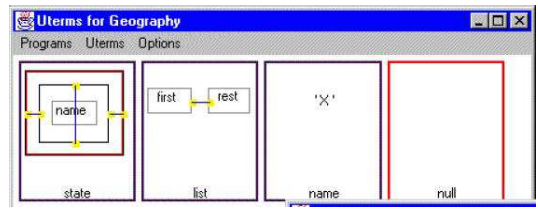}}
\caption{Example of query formulation in Delaunay}
\label{fig:delaunay}
\end{figure}

The Delaunay query language interface~\cite{DBLP:dblp_journals/vlc/CruzL01}   supports
standard SQL clauses including select, from and where. The queries are specified
on a class and the way of working is that the user arranges graphical objects
that specify how to visualize objects of the class. The complete specification of
a visual query is called DOODLE program. In Figure~\ref{fig:delaunay}, we
can see a DOODLE program which specifies a bar chart visualization.

\begin{figure}
\centerline{\includegraphics[width=60mm]{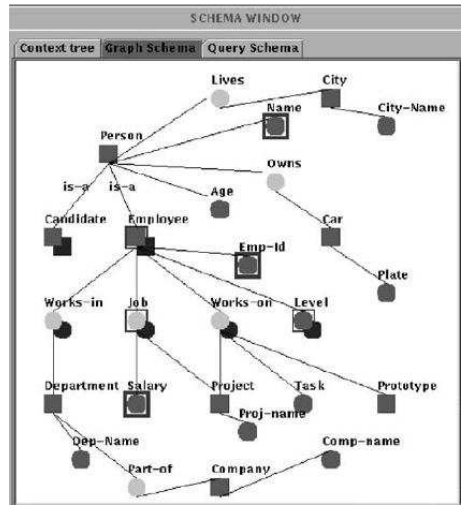}}
\caption{Example of visual query in TVQE.}
\label{fig:tvqe}
\end{figure}

In paper~\cite{DBLP:dblp_journals/is/SilvaCS02}, 
the queries are specified on the notebook, which visualizes schemas.
The approach is hybrid because
it is a combination of diagrammatic and iconic representation.
The query condition is specified by means of dialog boxes, which contain
menus, spin boxes and sliders for specifying temporal constants, predicates
on time instants or predicates on time periods. 
For example, the query 
`which salaries did the employees earn when they changed their level
for the last time' is represented in this system as in Figure~\ref{fig:tvqe}, 
where we can see several selected and displayed nodes in the query.

\begin{figure}
\centerline{\includegraphics[width=60mm]{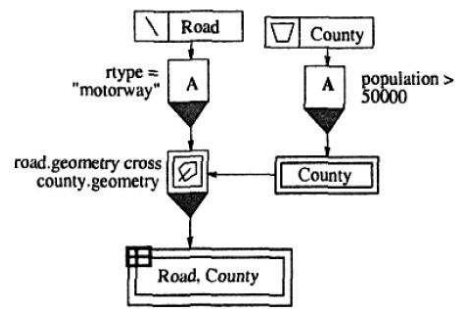}}
\caption{Example of query formulation in a GIS.}
\label{fig:gis1}
\end{figure}

Paper~\cite{Morris:2002:VQL:1556262.1556321} concentrates on
specifying queries for Geographical Information Systems. 
Query diagrams are constructed using filters, represented by icons, between data
input and output elements. The filter indicates the type of constraint expressed.
Two types of join operations are possible: nonspatial joins and spatial joins,
the latter are expressions of spatial relationships between spatial objects.
In Figure~\ref{fig:gis1} we can see the query `find all the motorway roads that cross
counties with population more than 50000'

\begin{figure}
\centerline{\includegraphics[width=60mm]{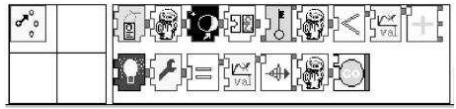}}
\caption{Example of query formulation in Iconic SQL}
\label{fig:iconic1}
\end{figure}

In paper
~\cite{DBLP:dblp_journals/vlc/NarayananS02},
the SQL sentences include only
SELECT, FROM, WHERE and ORDER BY clauses.
Queries are specified in a Query Assembly Frame, which consists of
two subframes: the Query Header Frame (QHF) and the Selection/
Search Criteria Frame (SCF). 
The QHF accommodates the SELECT verb and its attributes
as a composite icon. The SCF receives and accommodates 
sequences of icons representing the Selection/Search criteria.
For example the query 
`find the name, age and monthly salary of employees whose
age is less than a particular value and has a particular skill. Order
the result by age'
is expressed as in Figure~\ref{fig:iconic1}.

\begin{figure}
\centerline{\includegraphics[width=100mm]{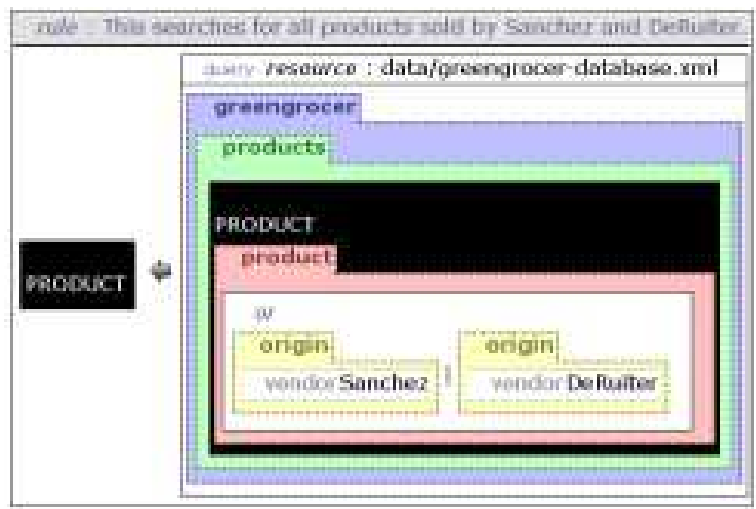}}
\caption{Example of query formulation in visXcerpt}
\label{fig:visxcerpt}
\end{figure}

The language visXcerpt~\cite{berger2003xcerpt}
uses a
pattern-based approach
to query 
XML data.
It is 
well suited for visual language because
the patterns are two dimensional
structures that conceptually are very close to two dimensional
visual representations. An example of a query in visXcerpt
is `select product elements having an origin with vendor attribute Sanchez or origin with vendor attribute DeRuiter'.
Its graphical representation can be seen in Figure~\ref{fig:visxcerpt}.

\begin{figure}
\centerline{\includegraphics[width=60mm]{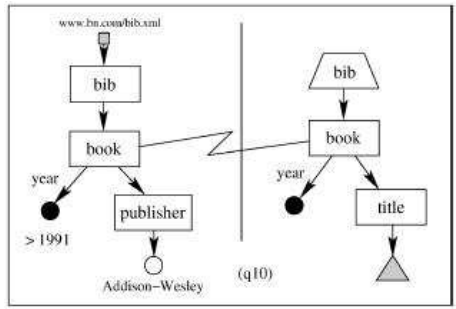}}
\caption{Example of visual Xquery in XQBE.}
\label{fig:xqbe}
\end{figure}

Paper~\cite{DBLP:journals/tods/BragaCC05} presents the XQBE visual query language
for expressing a large set of XQuery in a visual form. 
The schema-driven editing mode allows the user to build a query by means of a construction
guided by available XML schema or DTD specifications. The user draws the queries by choosing
the graphical constructs from the toolbar. For example, the XQBE query 
`list books published by Addison-Wesley after 1991, including their year and title'
can be seen in
Figure~\ref{fig:xqbe}.

\begin{figure}
\centerline{\includegraphics[width=60mm]{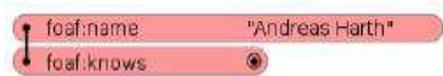}}
\caption{Query on RDF data}
\label{fig:rdf}
\end{figure}

The paper~\cite{DBLP:dblp_conf/www/HarthKD06} deals with the graphical representation of queries over semistructured data
on the web represented as RDF, a W3C standard. 
They define a facet, which can be seen as a filter condition over an RDF graph and offer a graphical representation of it. 
For example, in Figure~\ref{fig:rdf} we can see the query
`get resources that Andreas Harth knows'

\begin{figure}
\centerline{\includegraphics[width=60mm]{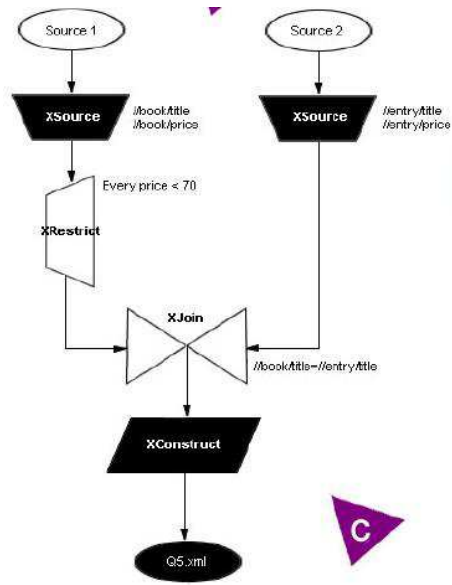}}
\caption{A query with join in IFOX}
\label{fig:ifox}
\end{figure}

In~\cite{DBLP:dblp_conf/sac/SansL08}, an interface for ordered XQuery, called IFOX, is presented.
The queries are represented as trees 
where the leaves are XSource operators,
the internal nodes are XOperators such as XJoin or XUnion
and the root node is the
XOperator XConstruct. 
For example, the query `For each book found
in http://localhost/bib.xml whose price is below 70 and where
the book is also found in http://localhost/review.xml, list the title of
the book and its price from each source' is expressed in IFOX as shown in Figure~\ref{fig:ifox}.
The query includes the operators XSource, Xjoin, XRestrict and XConstruct.

\begin{figure}
\centerline{\includegraphics[width=90mm]{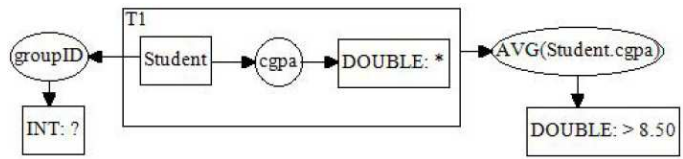}}
\caption{A query in CGBDI.}
\label{fig:querycgbdi}
\end{figure}

In CGBDI~\cite{varga2010conceptual}, the queries are specified also as a 
simple conceptual graph. For example, the query 
`select for every student group,
the average of student`s cgpa, if the average is greater than or
equal to 8.5'
is represented as in Figure~\ref{fig:querycgbdi}

\begin{figure}
\centerline{\includegraphics[width=60mm]{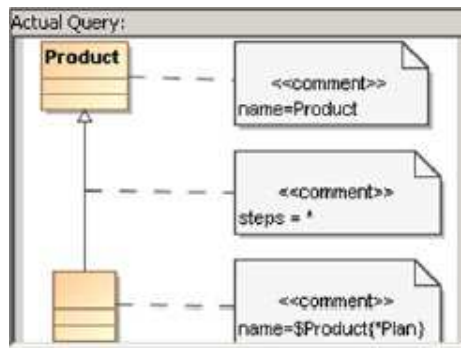}}
\caption{A query in VMQL.}
\label{fig:queryvmql}
\end{figure}

Queries in VMQL~\cite{storrle2011vmql} are drawn by using a palette of icons which offers
the icons of the source metamodel. So, there appears icons like Class, Package
or Generalization. The query can be seen in Figure~\ref{fig:queryvmql}.

\begin{figure}
\centerline{\includegraphics[width=60mm]{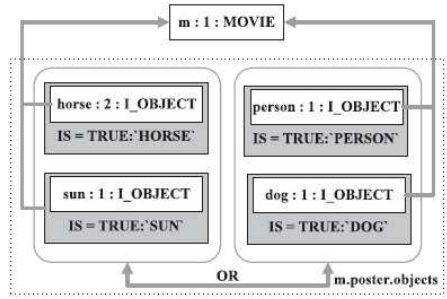}}
\caption{A query in GMQL.}
\label{fig:gmql}
\end{figure}

In GMQL~\cite{wu2012gmql}, queries attempt to find
information from multimedia databases. 
They have chosen a reduced but meaningful set
of symbols to express queries. An example can be
seen in Figure~\ref{fig:gmql} 
where,
the movies satisfying that: (1) the poster of each
one of these movies contains three inner image objects: two
`HORSE' and one `SUN'; or (2) the poster of each one of these movies
contains two inner image objects: one `PERSON' and one `DOG'
are found.

\subsubsection{Form-based}

Paper~\cite{sengupta1997query} generalizes QBE for databases
containing complex structured data. 
A QBT interface displays a template
and the user issues a query by entering ex-
amples of what (s)he is searching for in the template.

In~\cite{cohen1999equix} the Equix system is presented. 
A query form is created in HTML from the DTD of the underlying
database. Intuitively, the user creates an example of the
document (s)he is searching for.

\begin{figure}
\centerline{\includegraphics[width=120mm]{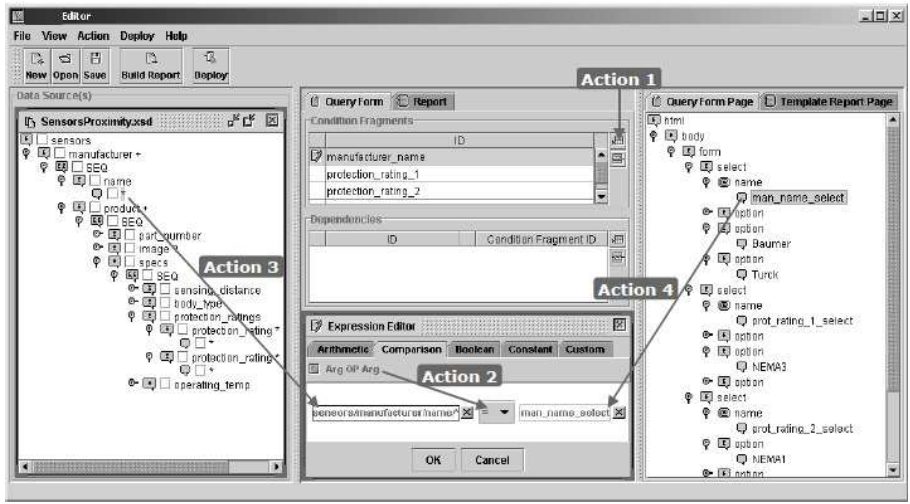}}
\caption{Query interface in QURSED.}
\label{fig:qursed}
\end{figure}

Paper~\cite{papakonstantinou2002qursed} presents QURSED
which is a visual query system generator rather than only a visual
query system. The Editor takes as input an
XML Schema and generates an interface
for the development of web-based query forms and
reports (QFRs). An example can be seen in Figure~\ref{fig:qursed}.

\begin{figure}
\centerline{\includegraphics[width=80mm]{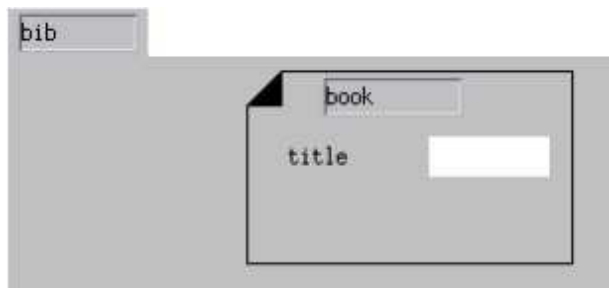}}
\caption{Query interface in FoXQ.}
\label{fig:foxq}
\end{figure}

FoXQ~\cite{abraham2003foxq}, based on~\cite{erwig2003xing}, 
presents a system that helps 
users build queries on XML documents incrementally by
navigating through layers of forms. Next, the queries are translated into XQuery.
For example, the query `return a list of all the titles
in a bibliography database' is expressed as in Figure~\ref{fig:foxq}.

\begin{figure}[t]
\centerline{\includegraphics[width=120mm]{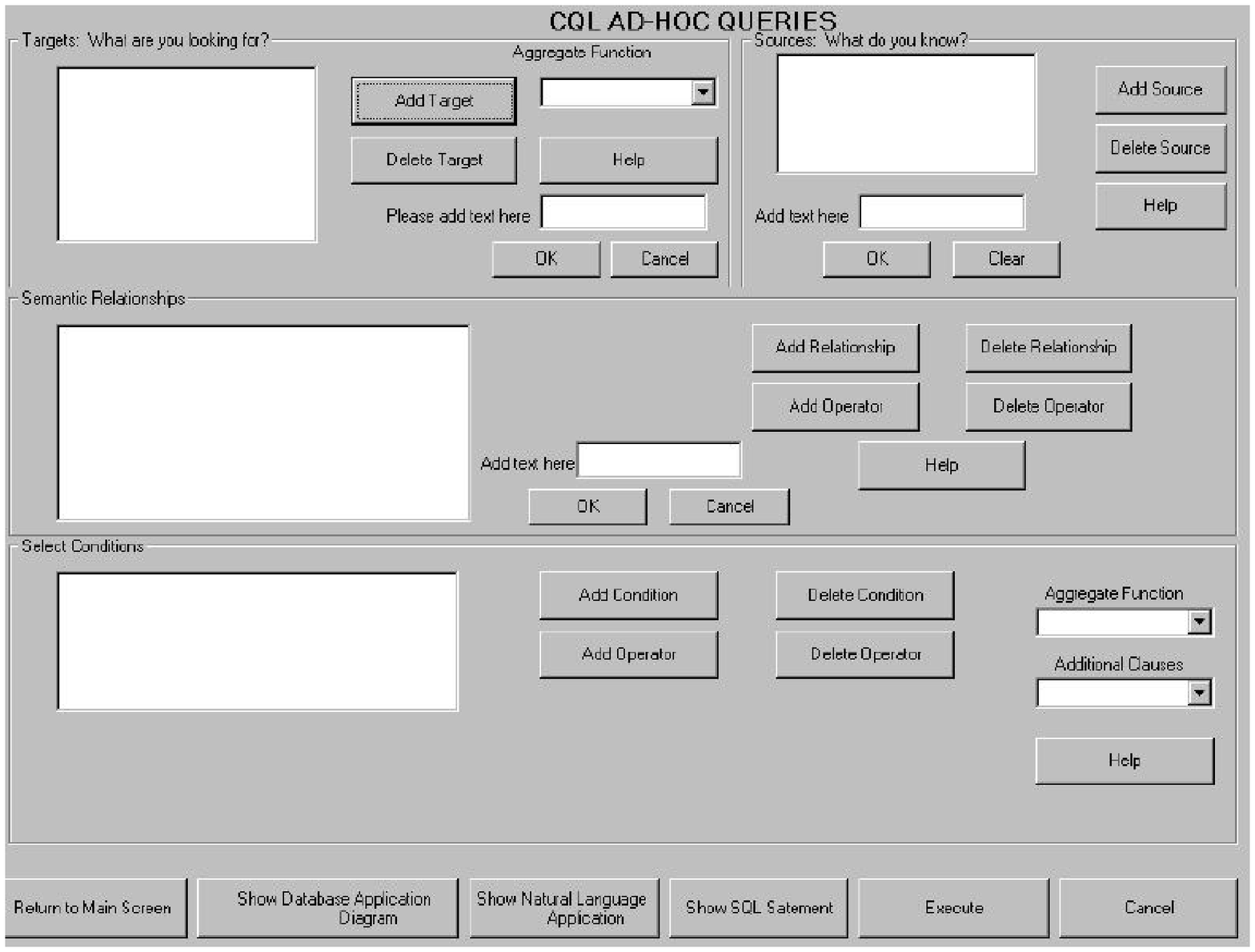}}
\caption{Query interface in CQL.}
\label{fig:cql}
\end{figure}

With the aid of users, in paper~\cite{DBLP:journals/cj/Owei03} a form-based interface
for querying databases was developed (see Figure~\ref{fig:cql}). This interface followed
well-established guidelines on the design of human interfaces. It has four clearly 
distinguished zones: targets or information you want to know, sources or information
you have available, semantic relationships and select conditions.
%%CQL is relationally complete, that is, it is equivalent in expressive power to the relational algebra.
%%tendría que encontrar un ejemplo de consulta en CQL

Paper~\cite{DBLP:dblp_conf/sigmod/JagadishCEJLNY07} proposes up to five interfaces
to query the MiMI system. One of them is the forms interface, from which the authors
discuss the pros and cons. The main pro is that it does not require knowledge of query language
for novice users. The main con is that the limited number of queries available in the interface might 
`dissatisfy many users'.

Paper~\cite{DBLP:dblp_journals/dke/TerwilligerDL07} presents the GUi As View (Guava) framework
for generating query interfaces from user interfaces for domain experts with little technical skills
to understand query data.
The rationale behind this is as follows: The interface for data capturing is typically designes to be
easy to use by domain experts. However, if they want to query data, there are two options: (a) have
someone write a special query interface or (b) use SQL to express queries against the database schema. 
The proposal of the paper is to follow option (a) but the query interface is generated automatically from the
user interface by  the Guava framework.
The approach of Guava
is equivalent in expressive power 
to single-statement conjunctive queries 
where joins are restricted to foreign keys and 
selection can use any of the six comparators to relate a
column against a constant, but not a column against another column.

\begin{figure}
\centerline{\includegraphics[width=80mm]{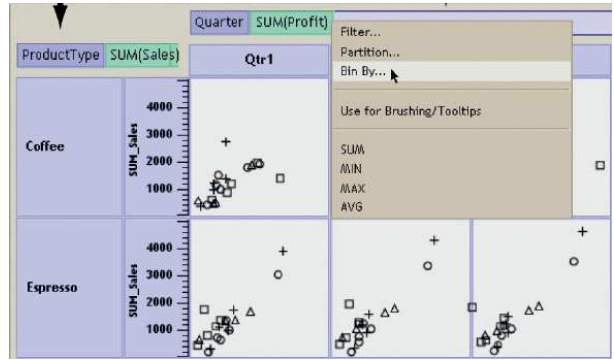}}
\caption{Query interface in Polaris.}
\label{fig:polaris}
\end{figure}

In Polaris~\cite{stolte2008polaris}, the user specifies a query by means of a table,
which consists of a number of rows,
columns, and layers. In the rows and columns of the table,
the user specifies fields of the database as well as
data aggregations. Each table entry, or pane, contains a set
of records, built from these data, 
that are visually encoded as a set of marks to create a graphic.
For example, a query for showing sales versus profit for different product types in different
quarters can be seen in Figure~\ref{fig:polaris}. The user has specifed in the vertical axis
the product type and the total of sales for the product. In the horizontal axis, the user
has specified the quarter and the total profit of the sales.

%%With respect to the expressive power, the paper states that `any query expressible
%%in SQL can be expressed as a specification in the Polaris formalism'.

\subsubsection{Faceted}
\begin{figure}
\centerline{\includegraphics[width=100mm]{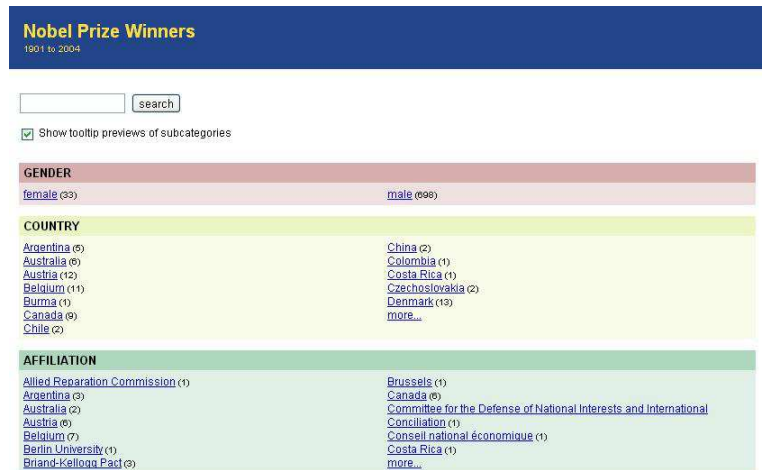}}
\caption{Faceted interface Flamenco.}
\label{fig:flamenco}
\end{figure}

Paper~\cite{DBLP:dblp_conf/jcdl/ClarksonNF09} presents faceted user interfaces,
which operate on faceted metadata by letting users select one or more values
from any number of facets. The advantage for users is that the user interface 
can present only valid selections, but this simplification also has drawbacks:
the expressivity is sacrificed by only allowing certain classes of queries, for example,
only conjunctive conditions. An example of a faceted interface is Flamenco
(see Figure~\ref{fig:flamenco}) where the items are Nobel Prizes.

%%FIN DE VISUAL REPRESENTATION ADOPT TO EXPRESS THE QUERIES

\subsection{Web orientation}

In column 4 of Table~\ref{tab:features1} and of Table~\ref{tab:features2},  we summarise the orientation to the web of the VQSs.

Regarding the data format, there are 9 papers
~\cite{cohen1999equix,papakonstantinou2002qursed,berger2003xcerpt,abraham2003foxq,DBLP:journals/tods/BragaCC05,meuss2005visual,DBLP:dblp_conf/sigmod/JagadishCEJLNY07,
DBLP:dblp_conf/sac/SansL08,choi2014vxq}
out of 34 which query XML data
and only two papers which query RDF data ~\cite{DBLP:dblp_conf/www/HarthKD06,DBLP:dblp_journals/internet/JarrarD10}. The rest of the papers do not query 
web data.

The papers were available on the web~\cite{DBLP:dblp_journals/vlc/CruzL01,DBLP:journals/tods/BragaCC05,stolte2008polaris,DBLP:dblp_conf/jcdl/ClarksonNF09,borges2010feasible}

Delaunay~\cite{DBLP:dblp_journals/vlc/CruzL01} offers a prototype implemented in Java and is available on the www.

XQBE~\cite{DBLP:journals/tods/BragaCC05} XQBE is fully implemented in a tool environment published on the Web
and the implementation is based on a client-server
architecture.

The POLARIS system~\cite{stolte2008polaris} has been adapted to the web,
so it is possible to perform analysis within a browser.

With the idea of the faceted interface, paper~\cite{DBLP:dblp_conf/jcdl/ClarksonNF09} points out that
several prototypes are available on the web, for example, the prototype Flamenco from the University of Berkeley.

In~\cite{borges2010feasible}, the   VISQUE system is presented. This is based on a client-server
architecture that includes a front-end which is
a web-based authoring tool that allows users to visually
interact with data elements and compose queries by
example.

\subsection{Validation}

In column 5 of Table~\ref{tab:features1} and of Table~\ref{tab:features2},  we summarise the validation method of the VQSs.

\subsubsection{Only prototype}

There is a prototype of VISUAL~\cite{DBLP:conf/icde/BalkirSOO96} which implements
a mapping from VISUAL to OQL. The paper claims that it has been designed with efficiency and ease of use.

Paper~\cite{DBLP:dblp_conf/er/BloeschH97} presents a query engine for
Conquer-II  written in C++. It consists of about 25000 lines of code and 83000 lines
shared with InfoModeler.

Paper~\cite{DBLP:conf/ideas/MadurapperumaGF97} has implemented a prototype where the user can select between
OMT, EER and a tabular representation for displaying schemas. On top of this, 
the query prototype is capable of handling select/project/join queries.

Paper~\cite{DBLP:journals/tkde/PoulovassilisH01} implements an architecture of
Hyperlog, which consists of several interconnected modules. The user interacts with
the editor for expressing queries.

%%The paper~\cite{DBLP:dblp_conf/www/HarthKD06} intends to verify 
%%their notation by implementing a user interface and to include in the system client-side scripts
%%to be able to interactively compose queries within a web browser.

Ifox~\cite{DBLP:dblp_conf/sac/SansL08} offers a prototype with five areas(see Figure~\ref{fig:ifox}):
input visualization, query visualization, output visualization, operator selection and 
operator specification. The first area show the XML schema, the second is used for
graphically specifying the query. The third shows the result of the query. The last two
are about the  Xoperators of the queries.

\subsubsection{Prototype tested with users}

Paper~\cite{sengupta1997query}  presents QBT and was tested with
novice and experienced users. The results show that QBT
is significantly more satisfying to
the users than traditional form-based approaches.

Paper~\cite{DBLP:dblp_journals/vldb/CatarciSC97} offers a multiparadigm interface
`where various equivalent
user interfaces can be defined on the basis of the same
formalisms, and different databases can be dealt with in a
uniform way'. They state that this interface was being tested with
real users but they do not offer any result of this testing.

The VISIONARY system~\cite{DBLP:dblp_journals/vlc/BenziMR99} offers
a section of comparison experiments with QBD*~\cite{DBLP:dblp_journals/tse/AngelaccioCS90} 
and QBI~\cite{378749}. As a conclusion,
VISIONARY is slightly better than QBD* and 
much better than QBI.

Paper~\cite{Morris:2002:VQL:1556262.1556321} implements a prototype with
the DELPHI system. Evaluation is underway with two categories of users:
with experience and without experience in GIS.

Iconic SQL~\cite{DBLP:dblp_journals/vlc/NarayananS02} offers a fully functional and working
implementation developed under Linux on a Pentium PC. They performed 
an evaluation to see how the context influences identification and association.
For this purpose, they set people a task and asked them to pick out relevant icons.
In some cases, the user did not guess correctly the icons but the more the
context develops in a session, the fewer details are needed in an icon for
identification purposes.

The initial interface of TVQE~\cite{DBLP:dblp_journals/is/SilvaCS02} 
was improved based on
the suggestions of users who had participated in a
preliminary experiment with the TVQE prototype, giving place to the version
described in the paper. The prototype was implemented
in Java using JDBC as interface to the DBMS.

The development of CQL~\cite{DBLP:journals/cj/Owei03} adopted a user-centered approach
in that it `involves participants who are representative of the intended
class of end-users in the interface development process'.

The first step was the elaboration of a mock-up with the purpose of eliciting
comments from end-users. These comments led to a second version of
the interface. The two interfaces were implemented as prototypes and were compared.
The conclusions are that end-users perform better with the second prototype and
they take lesss time to formulate queries with the first prototype.
Taking these results into account, they prepared the final production interface 
shown in Figure~\ref{fig:cql}.

In~\cite{borges2010feasible}, the authors issued a study for obtaining users' satisfaction related
to the ease of use of VISQUE, which reflects that the tool is easy to use and that 
casual users may need some initial help with the tool. The study was done with 
12 real users having domain knowledge about database systems but not about SQL.

The idea of Gblender~\cite{jin2010gblender} was tested with three users and was centered on the response
time rather than on the usability. With respect to the response time, GBlender outperforms
other state-of-the-art graph proposals.

The VMQL system~\cite{storrle2011vmql} has been tested with several users 
in order to establish the degree of usability
as compared to OCL and the logical query facility.
The results show better results for all tasks done by VMQL than
done by OCL. The tasks were reading queries or writing queries.

\subsubsection{Tested in a real environment}

Paper~\cite{DBLP:dblp_journals/dke/TerwilligerDL07} has been validated by the use of 
CORI analysts. CORI seeks to improve the practice of endoscopy by conducting studies 
on patient data. They distribute a software with which the clinician 
enters data that describes endoscopic procedures. From this data,
the analysts elaborate reports for use by physicians.
The Guava framework helps in extracting the data from the database.

Paper~\cite{DBLP:dblp_conf/avi/RontuKM06} proposes the SEEQ system, which
is utilized in the student information system of the university of the authors. The tool
is implemented in Java, uses the Swing GUI toolkit and the connexion to the database
is by means of JDBC.

The MiMi system~\cite{DBLP:dblp_conf/sigmod/JagadishCEJLNY07} proposes up to five interfaces
and has tested them with biologists. They draw many interesting conclusions. For example,
that ``when users are presented with multiple ways to access the information but do not
understand the underlying differences between the views, they tend to become confused
and lose trust in the system''.
Moreover, they have identified the problems non-expert users have with the
interfaces. Some of the problems are painful relations, painful options or unexpected pain.

\subsubsection{Commercial tool}

The POLARIS system of ~\cite{stolte2008polaris} has been widely tested because
it has been commercialized by Tableau Software as Tableau Desktop and is currently in use
by thousands of companies and tens of thousands of users. With all this experience,
they have validated the effectiveness of their proposal.

\section{Discussion}

The arrival of the web brought with it more facilities for users to query databases.
As a consequence, users expect to access easily through the web databases situated
anywhere in the world.

For expert users, one solution is to express queries in query languages such as SQL or XQuery. However, for novice
users whose main concern is to extract data from the database but not the query languages 
themselves, learning SQL or XQuery is a huge task that is very far from their main concern.

One solution for novice users is to hide the complexity of query languages behind a visual scenary where
it is supposed that the complexity is softened with the aid of visual metaphors.
This is the idea of Visual Query Systems (VQS) defined in~\cite{DBLP:dblp_journals/vlc/CatarciCLB97} as
``systems for querying databases that use a visual representation to
depict the domain of interest and express related requests".

In this paper, we have reviewed basic features of Visual Query Systems,
such as the representation of databases and the representation of queries.
We have also considered the feature of accessing data formatted for the web.
Finally, we have reviewed two features we consider relevant to determine
whether the VQSs ease querying for  novel users: web availability and validation.
Next, we discuss the results for each of these features.

The majority of papers offer a diagrammatic representation of the database,
only four papers an iconic one
~\cite{DBLP:conf/icde/BalkirSOO96,DBLP:dblp_journals/vlc/CruzL01,DBLP:dblp_journals/is/SilvaCS02,DBLP:dblp_journals/vlc/NarayananS02}
and one paper with form representation~\cite{stolte2008polaris}.
For several reasons, there are many papers whose database representation is unknown.
For example, paper~\cite{DBLP:journals/cj/Owei03} hides the database and tries to guess the paths for
the query from the entities chosen by the user.

With respect to the query representation, the distribution is more balanced between the icon(12 papers),
the diagram(11 papers) and the form (8 papers) representation.
A special form of query, the faceted one, appears only in one paper~\cite{DBLP:dblp_conf/jcdl/ClarksonNF09}.

%%Regarding to the data format
%%%COmpletar aquí un párrafo sobre XML y RDF data
Regarding the data format, there are 9 papers
~\cite{cohen1999equix,papakonstantinou2002qursed,berger2003xcerpt,abraham2003foxq,DBLP:journals/tods/BragaCC05,meuss2005visual,DBLP:dblp_conf/sigmod/JagadishCEJLNY07,
DBLP:dblp_conf/sac/SansL08,choi2014vxq}
out of 34 which query XML data
and only two papers which query RDF data ~\cite{DBLP:dblp_conf/www/HarthKD06,DBLP:dblp_journals/internet/JarrarD10}. The rest of the papers do not query 
web data.

The rest of the features we have identified deal with the main question we have formulated in this paper, that is,
to what extent have  the VQS been the solution for novel users for querying databases?

For answering this question with respect to the web availability, we can distinguish two periods. From 1997 to 2003 (see Table~\ref{tab:features1}),
when the web usage was beginning to spread, there was only one paper
oriented to the web~\cite{DBLP:dblp_journals/vlc/CruzL01}.
This was very understandable because of the time needed
for reorienting the research into the new web setting.
In the period 2004 to 2014, 
only papers
~\cite{DBLP:journals/tods/BragaCC05,stolte2008polaris,DBLP:dblp_conf/jcdl/ClarksonNF09,borges2010feasible}
propose a web implementation (see Table~\ref{tab:features2}).
Although the number of web oriented papers in this period is
greater than in the 1997-2003 period, the low number of papers
indicates that web orientation has scarcely been taken into account.

%%contar cuántos artículos hay en cada categoría
For  the validation feature,  
we have found a great number of papers which have only a prototype or have been tested with users in reduced experiments.
Only three prototypes have been tested in 
real environments
~\cite{DBLP:dblp_conf/avi/RontuKM06,DBLP:dblp_conf/sigmod/JagadishCEJLNY07,DBLP:dblp_journals/dke/TerwilligerDL07} 
and we have found only one  commercial tool
~\cite{stolte2008polaris}.
So, few papers go beyond testing the prototype with a few users.

As a conclusion of these two features, very few papers are web oriented
and also very few papers offer a prototype which has been tested in a real environment.
In fact, the combination of both features is only found in paper~\cite{stolte2008polaris}.
Then, although the visual query systems seem to be
a great idea for easing the query process for  novice users, the reality is that very few papers
describe real implementations.

So, the answer to the main question of the paper is that, for the moment, VQSs have not
been a widely accepted solution for novel users. From this observation a new, more general question
arises

is there any solution for easing the specification of queries?

If the answer is no, novel users have to learn by themselves query languages or
they have to ask computer experts for the specification of queries. In the latter case,
no new research would be needed in this field. If the answer is `we do not know', then
new research is required in order to find simple visual query languages which help
novice users.

We strongly believe that the idea of VQSs is a good one and that the research should continue
in this direction. Recent papers such as~\cite{li2012usability} also support the  idea that
a solution for naive users is not available but is necessary in this world in which
the use of databases is democratized. The paper proposes as a solution visual systems
in which the user writes examples of queries and the system extracts and specifies
the desired query in the corresponding query language.

\bibliography{vqssurvey}{}
\bibliographystyle{plain}

\end{document}